\documentclass[twocolumn,prl,showpacs,superscriptaddress,preprintnumbers,amsmath,amssymb,floatfix]{revtex4}
\usepackage[dvips]{graphicx}
\usepackage{dcolumn}
\usepackage{bm}
\begin{document}
\title{The spin state transition in LaCoO$_{3}$; revising a revision}

\author{M.~W.~Haverkort}
  \affiliation{II. Physikalisches Institut, Universit\"{a}t zu K\"{o}ln, Z\"{u}lpicher Str. 77, 50937 K\"{o}ln, Germany}
\author{Z.~Hu}
  \affiliation{II. Physikalisches Institut, Universit\"{a}t zu K\"{o}ln, Z\"{u}lpicher Str. 77, 50937 K\"{o}ln, Germany}
\author{J.~C.~Cezar}
  \affiliation{European Synchrotron Radiation Facility, Bo\^ite Postale 220, 38043 Grenoble C\'edex, France}
\author{T.~Burnus}
  \affiliation{II. Physikalisches Institut, Universit\"{a}t zu K\"{o}ln, Z\"{u}lpicher Str. 77, 50937 K\"{o}ln, Germany}
\author{H.~Hartmann}
  \affiliation{II. Physikalisches Institut, Universit\"{a}t zu K\"{o}ln, Z\"{u}lpicher Str. 77, 50937 K\"{o}ln, Germany}
\author{M.~Reuther}
  \affiliation{II. Physikalisches Institut, Universit\"{a}t zu K\"{o}ln, Z\"{u}lpicher Str. 77, 50937 K\"{o}ln, Germany}
\author{C.~Zobel}
  \affiliation{II. Physikalisches Institut, Universit\"{a}t zu K\"{o}ln, Z\"{u}lpicher Str. 77, 50937 K\"{o}ln, Germany}
\author{T.~Lorenz}
  \affiliation{II. Physikalisches Institut, Universit\"{a}t zu K\"{o}ln, Z\"{u}lpicher Str. 77, 50937 K\"{o}ln, Germany}
\author{A.~Tanaka}
  \affiliation{Department of Quantum Matter, ADSM, Hiroshima University, Higashi-Hiroshima 739-8530, Japan}
\author{N.~B.~Brookes}
  \affiliation{European Synchrotron Radiation Facility, Bo\^ite Postale 220, 38043 Grenoble C\'edex, France}
\author{H.~H.~Hsieh}
  \affiliation{Chung Cheng Institute of Technology, National Defense University, Taoyuan 335, Taiwan}
  \affiliation{National Synchrotron Radiation Research Center, 101 Hsin-Ann Road, Hsinchu 30076, Taiwan}
\author{H.-J.~Lin}
  \affiliation{National Synchrotron Radiation Research Center, 101 Hsin-Ann Road, Hsinchu 30076, Taiwan}
\author{C.~T.~Chen}
  \affiliation{National Synchrotron Radiation Research Center, 101 Hsin-Ann Road, Hsinchu 30076, Taiwan}
\author{L.~H.~Tjeng}
  \affiliation{II. Physikalisches Institut, Universit\"{a}t zu K\"{o}ln, Z\"{u}lpicher Str. 77, 50937 K\"{o}ln, Germany}

\date{\today}

\begin{abstract}
Using soft x-ray absorption spectroscopy and magnetic circular
dichroism at the Co-$L_{2,3}$ edge we reveal that the spin state
transition in LaCoO$_{3}$ can be well described by a low-spin
ground state and a triply-degenerate high-spin first excited
state. From the temperature dependence of the spectral lineshapes
we find that LaCoO$_{3}$ at finite temperatures is an
inhomogeneous mixed-spin-state system. Crucial is that the
magnetic circular dichroism signal in the paramagnetic state
carries a large orbital momentum. This directly shows that the
currently accepted low-/intermediate-spin picture is at variance.
Parameters derived from these spectroscopies fully explain
existing magnetic susceptibility, electron spin resonance and
inelastic neutron data.
\end{abstract}


\maketitle

LaCoO$_{3}$ shows a gradual non-magnetic to magnetic transition
with temperature, which has been interpreted originally four
decades ago as a gradual population of high spin (HS,
$t_{2g}^4e_{g}^2$, $S=2$) excited states starting from a low spin
(LS, $t_{2g}^6$, $S=0$) ground state
\cite{Heikes64,Blasse65,Naiman66,Jonker66,Goodenough65,Raccah67,Goodenough71,Bhide72}.
This interpretation continued to be the starting point for
experiments carried out up to roughly the first half of the 1990's
\cite{Asai94,Itoh94,Itoh95,Yamaguchi96}. All this changed with the
theoretical work in 1996 by Korotin \textit{et al.}, who proposed
on the basis of local density approximation + Hubbard U (LDA+U)
band structure calculations, that the excited states are of the
intermediate spin (IS, $t_{2g}^5e_{g}^1$, $S=1$) type
\cite{Korotin96}. Since then many more studies have been carried
out on LaCoO$_{3}$ with the majority of them
\cite{Saitoh97,Stolen97,Asai98,Okamoto00,Ravindran02,Zobel02,Radaelli02,Vogt03,Nekrasov03,Louca03,Maris03,Ishikawa04,Magnuson04,Knizek05}
claiming to have proven the presence of this IS mechanism. In
fact, this LDA+U work is so influential \cite{Korotincitation}
that it forms the basis of most explanations for the fascinating
properties of the recently synthesized layered cobaltate
materials, which show giant magneto resistance as well as
metal-insulator and ferro-ferri-antiferro-magnetic transitions
with various forms of charge, orbital and spin ordering
\cite{refHu04,Hu04}.

In this paper we critically re-examine the spin state issue in
LaCoO$_{3}$. There has been several attempts made since 1996 in
order to revive the LS-HS scenario
\cite{Zhuang98,Noguchi02,Kyomen03,Ropka03,Kyomen05}, but these
were overwhelmed by the above mentioned flurry of studies claiming the IS mechanism
\cite{Saitoh97,Stolen97,Asai98,Okamoto00,Ravindran02,Zobel02,Radaelli02,Vogt03,Nekrasov03,Louca03,Maris03,Ishikawa04,Magnuson04,Knizek05}.
Moreover, a new investigation using inelastic neutron scattering
(INS) has recently appeared in Phys. Rev. Lett. \cite{Phelan06}
making again the claim that the spin state transition involves the
IS states. Here we used soft x-ray absorption spectroscopy (XAS)
and magnetic circular dichroism (MCD) at the Co-$L_{2,3}$ edge
and we revealed that the spin state transition in LaCoO$_{3}$ can
be well described by a LS ground state and a triply degenerate HS
excited state, and that an inhomogeneous mixed-spin-state system
is formed. Parameters derived from these spectroscopies fully
explain existing magnetic susceptibility and electron spin
resonance (ESR) data, and provide support for an alternative
interpretation of the INS \cite{Podlesnyak05}. Consequently the
spin state issue for the new class of the layered cobaltates
needs to be reinvestigated \cite{refHu04,Hu04}.

Single crystals of LaCoO$_{3}$ have been grown by the traveling
floating-zone method in an image furnace. The magnetic
susceptibility was measured using a Quantum Design vibrating
sample magnetometer (VSM), reproducing the data reported earlier
\cite{Zobel02}. The Co-$L_{2,3}$ XAS measurements were performed
at the Dragon beamline of the National Synchrotron Radiation
Research Center (NSRRC) in Taiwan with an energy resolution of
0.3 eV. The MCD spectra were collected at the ID08 beamline of
the European Synchrotron Radiation Facility (ESRF) in Grenoble
with a resolution of 0.25 eV and a degree of circular
polarization close to $100\%$ in a magnetic field of 6 Tesla.
Clean sample areas were obtained by cleaving the crystals
\textit{in-situ} in chambers with base pressures in the low
10$^{-10}$ mbar range. The Co $L_{2,3}$ spectra were recorded
using the total electron yield method (TEY). O-$K$ XAS spectra
were collected by both the TEY and the bulk sensitive fluorescence
yield (FY) methods, and the close similarity of the spectra taken
with these two methods verifies that the TEY spectra are
representative for the bulk material. A CoO single crystal is measured \textit{simultaneously} in a
separate chamber to obtain relative energy referencing with
better than a few meV accuracy, sufficient to extract reliable
MCD spectra.

\begin{figure}[ht]
     \includegraphics[width=0.45\textwidth]{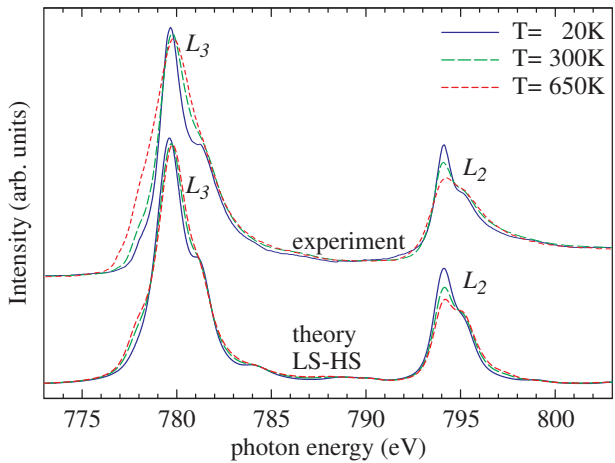}
     \caption{(color online) Experimental Co-$L_{2,3}$ XAS spectra taken from LaCoO$_{3}$
     at various temperatures between 20 and 650 K, together with the
     corresponding theoretical isotropic spectra calculated using a CoO$_{6}$
     cluster in the LS-HS scenario. For clarity, only the 20, 300 and 650 K
     spectra are shown.}
     \label{fig1}
\end{figure}

Fig. 1 shows the set of Co-$L_{2,3}$ XAS spectra of LaCoO$_{3}$
taken for a wide range of temperatures. The set is at first sight
similar to the one reported earlier \cite{Abbate93}, but it is in
fact essentially different in details. First of all, our set
includes a low temperature (20 K) spectrum representative for the
LS state, and second, our spectra do not show a
pronounced shoulder at 777 eV photon energy which is
characteristic for the presence of Co$^{2+}$ impurities
\cite{Csiszar05}. The extended temperature range and especially the purity
of the probed samples provide the required sensitivity for the
spin-state related spectral changes.

The spectra are dominated by the Co $2p$ core-hole spin-orbit
coupling which splits the spectrum roughly in two parts, namely
the $L_{3}$ ($h\nu \approx 780$ eV) and $L_{2}$ ($h\nu \approx
796$ eV) white lines regions. The line shape of the spectrum
depends strongly on the multiplet structure given by the Co
$3d$-$3d$ and $2p$-$3d$ Coulomb and exchange interactions, as
well as by the local crystal fields and the hybridization with
the O $2p$ ligands. Unique to soft x-ray absorption is that the
dipole selection rules are very effective in determining which of
the $2p^{5}3d^{n+1}$ final states can be reached and with what
intensity, starting from a particular $2p^{6}3d^{n}$ initial
state ($n$=6 for Co$^{3+}$) \cite{deGroot94,Thole97}. This makes
the technique extremely sensitive to the symmetry of the initial
state, e.g. the spin state of the Co$^{3+}$ \cite{Hu04}.

Utilizing this sensitivity, we first simulate the spectrum of a
Co$^{3+}$ ion in the LS state using the successful configuration
interaction cluster model that includes the full atomic multiplet
theory and the hybridization with the O $2p$ ligands
\cite{deGroot94,Thole97,Tanaka94}. The CoO$_6$ cluster is taken
to have the octahedral symmetry and the parameters are the same
as the ones which succesfully reproduce the spectrum of LS
EuCoO$_{3}$ \cite{Hu04,paramcluster}. The result with the ionic
part of the crystal field splitting set at $10Dq=0.7$ eV is shown
in Fig. 1 and fits well the experimental spectrum at 20 K.

\begin{figure}[ht]
     \includegraphics[width=0.45\textwidth]{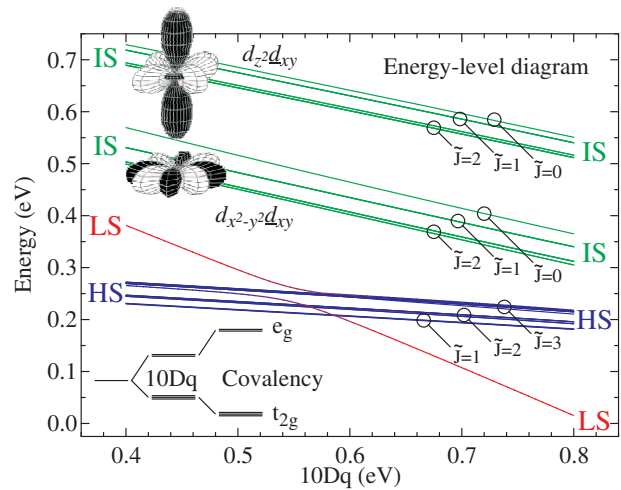}
     \caption{(color online) Energy level diagram of a CoO$_{6}$ cluster [43] as a function of
     the ionic part of the crystal field splitting $10Dq$.}
     \label{fig2}
\end{figure}

Next we analyze the spectra for the paramagnetic phase. We use
the same cluster keeping the $O_h'$ symmetry, and calculate the
total energy level diagram as a function of $10Dq$, see Fig. 2.
We find that the ground state of the cluster is either LS or HS
(and never IS) with a cross-over at about $10Dq=0.58$ eV \cite{10DqComment}. We are
able to obtain good simulations for the spectra at all
temperatures, see Fig. 1, provided that they are made from an
incoherent sum of the above mentioned LS cluster spectrum
calculated with $10Dq=0.7$ eV and a HS cluster spectrum calculated
with $10Dq=0.5$ eV. It is not possible to fit the entire
temperature range using one cluster with one particular
temperature-independent $10Dq$ value for which the ground state
is LS-like and the excited states HS-like. Moreover, each of
these two $10Dq$ values have to be sufficiently far away from the
LS-HS crossover point to ensure a large enough energy separation
between the LS and HS so that the two do not mix due to the
spin-orbit interaction. Otherwise, the calculated low temperature
spectrum, for instance, will disagree with the experimental one.
All this indicates that LaCoO$_{3}$ at finite temperatures is an
inhomogeneous mixed spin state system.

\begin{figure*}[ht]
     \includegraphics[width=1.0\textwidth]{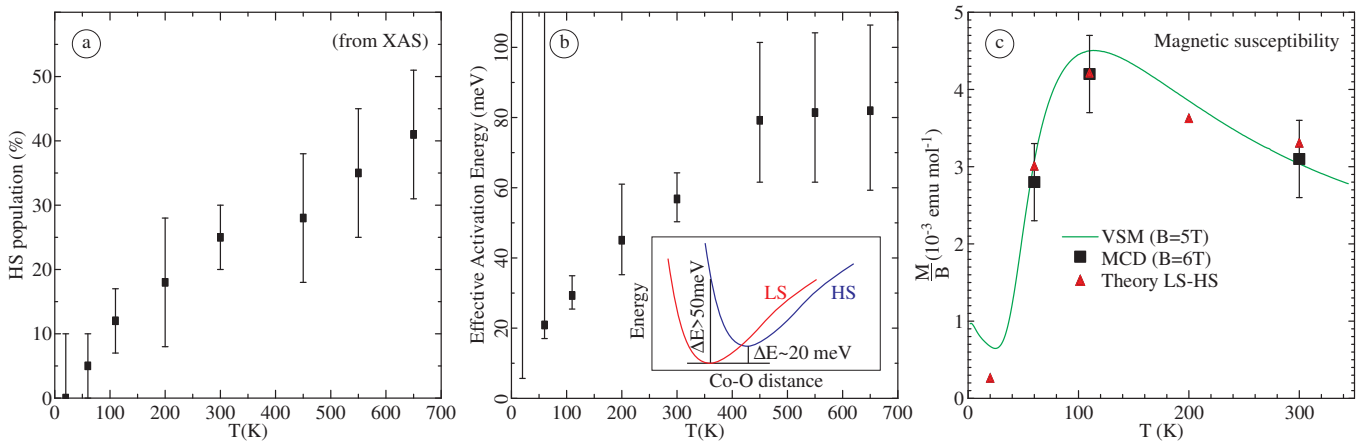}
     \caption{(color online) (a) The percentage of the HS population as obtained from the
     XAS data. (b) Corresponding effective activation energy between the LS
     and the lowest HS state. The inset sketches the role of lattice relaxations.
     (c) Magnetic susceptibility measured by VSM (solid line), calculated from the cluster
     (red triangles) using the HS population of Fig. 3a, and extracted from MCD data
     (black squares) of Fig. 4.}
     \label{fig3}
\end{figure*}

The temperature dependence has been fitted by taking different
ratios of LS and HS states contributing to the spectra. The
extracted HS percentage as a function of temperature is shown in
Fig. 3a. The corresponding effective activation energy is
plotted in Fig 3b. It increases with temperature and varies
between 20 meV at 20 K to 80 meV at 650 K,
supporting a recent theoretical analysis of the thermodynamics
\cite{Kyomen05}. Here we would like to point out that these
numbers are of the order $k_BT$ and reflect total energy
differences which include lattice relaxations \cite{Kyomen05} as
sketched in the inset of Fig. 3b. Without these relaxations, we
have for the LS state ($10Dq=0.7$ eV) an energy difference of at
least 50 meV between the LS and the HS as shown in Fig. 2. In
such a frozen lattice, the energy difference is larger than
$k_BT$. It is also so large that the ground state is indeed
highly pure LS as revealed by the 20 K spectrum.

To check the validity of our analysis, we calculate the magnetic
susceptibility using the CoO$_{6}$ cluster and the HS occupation
numbers from Fig. 3 as derived from the XAS data. The results are
plotted in Fig. 3c (red triangles) together with the magnetic
susceptibility as measured by the VSM (solid line). We can
observe clearly a very good agreement: the magnitude and its
temperature dependence is well reproduced. This provides another
support that the spin-state transition is inhomogeneous and
involves lattice relaxations. A homogeneous LS-HS model, on the
other hand, would produce a much too high susceptibility if it is
to peak at 110 K \cite{Itoh95,Yamaguchi96,Saitoh97,Zobel02}. In
addition, it is crucial to realize that the Van Vleck
contribution to the magnetic susceptibility strongly depends on
the intermixing between the LS and HS states. It is precisely
this aspect which also sets the condition that the energy
separation between the LS and HS states in the cluster should be
larger than 50 meV, otherwise the calculated Van Vleck
contribution would already exceed the experimentally determined
total magnetic susceptibility at low temperatures. This in fact
is a restatement of the above mentioned observation that the low
temperature spectrum is highly pure LS.

\begin{figure}[ht]
     \includegraphics[width=0.45\textwidth]{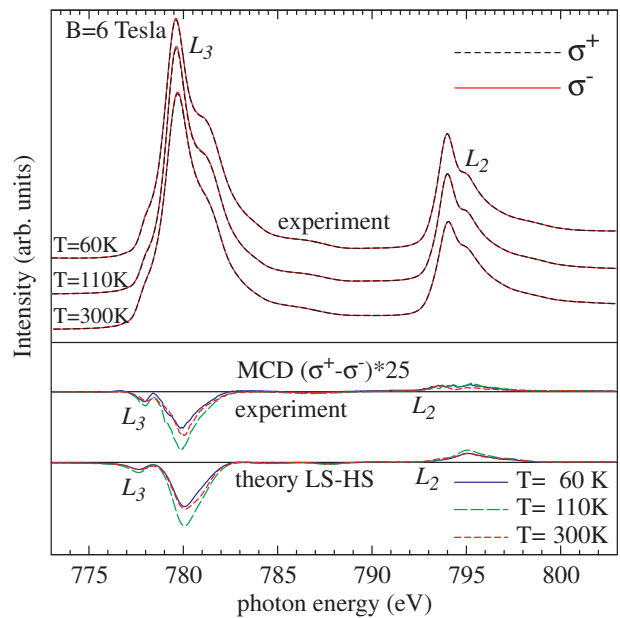}
     \caption{(color online) Top curves: experimental Co-$L_{2,3}$ XAS spectra taken from LaCoO$_{3}$
     at 60, 110, and 300 K using circularly polarized x-rays with the photon spin aligned parallel
     (black dotted line, $\sigma^{+}$) and antiparallel (red solid line, $\sigma_{-}$) to the 6 Tesla magnetic field.
     Middle curves: experimental MCD spectra defined as the difference between the two spin alignments.
     Bottom curves: theoretical MCD spectra calculated in the LS-HS scenario.}
     \label{fig4}
\end{figure}

To further verify the direct link between the spectroscopic and
the VSM magnetic susceptibility data, we carried out MCD
experiments on LaCoO$_{3}$ at 60, 110 and 300K, i.e. in the
paramagnetic phase, using a 6 Tesla magnet. Fig. 4 shows XAS
spectra taken with circularly polarized soft-x-rays with the
photon spin parallel and antiparallel
aligned to the magnetic field. The
difference in the spectra using these two alignments is only of
the order of 1\%, but can nevertheless be measured reliably due
to the good signal to noise ratio, stability of the beam, and the
accurate photon energy referencing. The difference curves are
drawn in the middle of Fig. 4 with a magnification of 25x. Hereby
we have subtracted a small signal due to the presence of about
1.5\% Co$^{2+}$ impurities. We also plotted the simulated MCD
spectra from the cluster model within the LS-HS scenario, and we
can clearly observe a very satisfying agreement with the
experiment. Alternatively, using the MCD sum-rules developed by
Thole and Carra et al. \cite{Thole92,Carra93}, we can extract
directly the orbital ($L_z$) and spin (2$S_z$) contributions to
the induced moments without the need to do detailed modeling
\cite{param}. This result normalized to the applied magnetic
field is plotted in Fig. 3c, and we can immediately observe the
close agreement with the VSM data.

An important aspect that emerges directly from the MCD
experiments, is the presence of a very large induced orbital
moment: we find that $L_{z}/S_{z} \approx 0.5$. This means that
the spin-orbit coupling (SOC) must be considered in evaluating
the degeneracies of the different levels, as is done for the
energy level diagram in Fig. 2. Let us discuss the consequences
for the HS state. We see that the 15-fold degenerate (3-fold
orbital and 5-fold spin) HS state is split by the SOC. A $t_{2g}$
electron has a pseudo orbital momentum of \textit{\~L}$=1$
\cite{AbragamBleaney70} which couples with the spin to a pseudo
total momentum of \textit{\~J}$=1,2$ or $3$. The \textit{\~J}$=1$
triplet is the lowest in energy and we find from our cluster that
this state has $L_z=0.6$ and $S_z=1.3$, in good agreement with the
experimental $L_{z}/S_{z} \approx 0.5$. Realizing that this state
is a triplet with a spin momentum ($S_{z}$) so close to 1, it is
no wonder that many studies incorrectly interpreted this state as
an IS state. Its expectation value for the spin ($\langle
S^2\rangle=S(S+1)$) is however very close to 6 and the formal
occupation numbers of the $d_{z^2}$ and the $d_{x^2-y^2}$
orbitals are both equal to 1. This state is clearly a HS state
and should not be confused with an IS state. We find a $g$-factor
of 3.2, in good agreement with the values found from ESR
\cite{Noguchi02,Ropka03} and INS data \cite{Podlesnyak05}.

We have shown so far that the spin state transition in
LaCoO$_{3}$ is in very good agreement with a LS - HS picture. The
question now remains if it could also be explained within a LS -
IS scenario. For that we first have to look what the IS actually
is. The IS state has one hole in the $t_{2g}$ shell and one
electron in the $e_{g}$ shell. Due to the strong orbital
dependent Coulomb interactions, the strong-Jahn-Teller states of
the type $d_{z^2}\underline{d}_{xy}$ and their $x,y,z$-cyclic
permutations have much higher energies than the weak-Jahn-Teller
$d_{x^2-y^2}\underline{d}_{xy}$ plus cyclic permutations. Here
the underline denotes a hole. See Fig. 2. These weak-Jahn-Teller
states indeed form the basis for the orbital ordering scheme as
proposed for the IS scenario by Korotin \textit{et al.}
\cite{Korotin96}. However, these real-space states do not carry a
large orbital momentum, and are therefore not compatible with the
values observed in the MCD measurements. Likewise, the strong
Jahn-Teller-like local distortions in the IS state proposed by
Maris \textit{et al.} \cite{Maris03} would lead to a quenching of
the orbital momentum. We therefore can conclude that the IS
scenarios proposed so far have to be rejected on the basis of our
MCD results. Moreover, an IS state would lead in general to a
much larger van Vleck magnetism than a HS state. This is related
to the fact that the LS state couples directly to the IS via the
SOC, while the HS is not. To comply with the measured low
temperature magnetic susceptibility, the energy difference
between the LS and IS has to be 150 meV at least, making it more
difficult to find a mechanism by which the maximum of the
susceptibility occurs at 110K. Finally, within the LS-IS
scenario, we were not able to find simulations which match the
experimental XAS and MCD spectra.

To summarize, we provide unique spectroscopic evidence that the
spin state transition in LaCoO$_{3}$ can be well described by a LS
ground state and a triply degenerate HS excited state, and that an
inhomogeneous mixed-spin-state system is formed. The large
orbital momentum revealed by the MCD measurements invalidates
existing LS-IS scenarios. A consistent picture has now been
achieved which also explains available magnetic susceptibility,
ESR and INS data.

We would like to thank Lucie Hamdan for her skillful technical
and organizational assistance in preparing the experiments. The
research in Cologne is supported by the Deutsche
Forschungsgemeinschaft through SFB 608.


\begin{thebibliography}{99}
\bibitem{param} The sumrules gives numbers for $L_z$ and
 $S_{z}+\frac{7}{2}T_{z}$; $T_z=0$ for the HS state as justified
 from cluster calculations; the number of $3d$ holes is about 3.5.
\end{thebibliography}
\end{document}